\documentclass[conference, a4paper]{IEEEtran}
\usepackage{url}

\usepackage[utf8]{inputenc}
\usepackage{hyperref}
\usepackage{amsmath}
\usepackage{graphicx}
\usepackage{diagbox}
\usepackage{cite}
\usepackage{enumerate}

\usepackage{color}

\IEEEoverridecommandlockouts

\begin{document}

\title{Easy-to-Use On-the-Fly Binary Program Acceleration on Many-Cores}

\author{\IEEEauthorblockN{Marvin Damschen*\IEEEcompsocitemizethanks{* This work was conducted while Marvin Damschen was student at the University of Paderborn}}
\IEEEauthorblockA{Karlsruhe Institute of Technology \\
marvin.damschen@kit.edu}
\and
\IEEEauthorblockN{Christian Plessl}
\IEEEauthorblockA{University of Paderborn\\
christian.plessl@uni-paderborn.de}}

\maketitle

\begin{abstract}

This paper introduces Binary Acceleration At Runtime (BAAR), an easy-to-use on-the-fly binary acceleration mechanism which aims to tackle the problem of enabling existent software to automatically utilize accelerators at runtime. BAAR is based on the LLVM Compiler Infrastructure and has a client-server architecture. The client runs the program to be accelerated in an environment which allows program analysis and profiling. Program parts which are identified as suitable for the available accelerator are exported and sent to the server. The server optimizes these program parts for the accelerator and provides RPC execution for the client. The client transforms its program to utilize accelerated execution on the server for offloaded program parts.
We evaluate our work with a proof-of-concept implementation of BAAR that uses an Intel Xeon Phi  5110P as the acceleration target and performs automatic offloading, parallelization and vectorization of suitable program parts. The practicality of BAAR for real-world examples is shown based on a study of stencil codes. Our results show a speedup of up to 4$\times$ without any developer-provided hints and 5.77$\times$ with hints over the same code compiled with the Intel Compiler at optimization level O2 and running on an Intel Xeon E5-2670 machine. Based on our insights gained during implementation and evaluation we outline future directions of research, e.g., offloading more fine-granular program parts than functions, a more sophisticated communication mechanism or introducing on-stack-replacement.

\end{abstract}

\IEEEpeerreviewmaketitle

\section{Introduction}
Modern computer systems supply multiple diverse computing resources in addition to standard instruction set single-core CPUs, such as multi and many-core processors and accelerators with with instruction set extensions, e.g.,  GPUS or Xeon Phi. To exploit the whole performance these computing resources provide, applications have to be optimized for the specific instruction set extensions and accelerators available. Manually adapting software to this ever-evolving technology is tedious, however. Automatic tools to alleviate this problem are needed. With the source code available, developers can make use of compiler suites which aid optimizing software to new technology. However, when dealing with closed-source binary software, such as commercial applications or libraries, developers cannot modify the code to exploit the capabilities of novel computing resources. The same problem arises also if the source code is available, but a lack of skills or programmers prevents an adaptation of the applications. Unfortunately, the possibilities to optimize compiled software for new technology are scarce. A widely used approach is just-in-time compilation, but it does not make use of accelerators and it is limited to the platform it is performed on. Therefore, a more general approach than just-in-time compilation is needed to exploit the full performance of modern computer systems when running existing software.

This work presents \emph{Binary Acceleration At Runtime (BAAR)}, an approach and implementation to enable easy-to-use on-the-fly binary program acceleration. BAAR has a client-server architecture. The BAAR Client provides an LLVM-based environment to execute binaries in the form of LLVM Immediate Representation (LLVM IR). Programs are profiled and analyzed in this environment in parallel to their execution.  Once enough information has been gathered to identify compute intensive parts of the program, the identified program parts are exported into a code fragment in LLVM IR format, which we denote as the \textit{remote part}. The remote part is sent to the BAAR Server which provides an acceleration target for the client. While the client makes progress running the original program, the server optimizes the remote part for the current acceleration target available. When the optimization finishes, the client is notified by the server. The client transforms the original program into the \textit{local part}, an extension of the original program with the possibility to offload calls to the remote part on the server with an remote procedure call mechanism. For each call to a function that has been exported to a remote part, the decision whether to execute the code locally or remote is decided at runtime depending on the function arguments.

Our approach does not require the source code of the programs to be available, as several projects have shown that it is feasible to transform binaries into LLVM IR~\cite{dagger,Anand:2013:CIR:2465351.2465380,Chipounov2011}. BAAR has been designed to be modular and flexible. For example, the communication mechanism between client and server is exchangeable, e.g., shared memory and TCP/IP connections have been tested. Furthermore, the client does not have to know the specific accelerator available on the server, only its characteristics; It is also conceivable for the server to provide multiple targets. Client architecture, communication mechanism and acceleration target are independently exchangeable, offering several potential use cases and optimization targets, e.g., improving performance of legacy code on a workstation or minimizing power consumption of low-power computing devices.

To evaluate the practicality of our approach, the BAAR prototype is used to execute two stencil applications taken from the Polybench \cite{polybench} benchmark suite. The client is run on an Intel Xeon E5-2670 machine, which communicates with the server over a simple TCP/IP-based protocol. The server is run on an Intel Xeon Phi accelerator card, which also functions as the target for automatic parallelization and vectorization and for executing the automatically offloaded function calls. Our measurements show that, without any developer-provided directives, a speedup of 4$\times$ is achievable when comparing the execution of the Jacobi 2D stencil with BAAR to the execution of native code generated with the Intel Compilers with optimization level O2. Further, our evaluation shows that an improved alias analysis would allow a speedup of 5.77$\times$ when executing the FDTD 2D stencil. As there is currently no Xeon Phi backend available in LLVM, we need to use a detour over C code generation and use of the Intel native compiler to obtain Xeon Phi binaries from LLVM IR. This impairs the quality of the vectorization and the performance of BAAR in general. BAAR is however prepared to utilize the native Xeon Phi LLVM backend once available, promising even higher speedups.

\section{Related Work}
Optimizing existent software for new technology is an active topic, several projects try to tackle this problem in various ways.

The Intel SPMD Program Compiler (ispc)~\cite{pharr2012ispc} follows a static approach. It compiles a C-based programming language with single program multiple data (SPMD) extensions, enabling programmers to exploit vector units and multiple CPU cores in a familiar language without the need to know intrinsics. For ispc, the program source code has to be available and adapted. Once compiled, there is no way of adapting the binary program to changing environments.

KernelGen~\cite{13kernelgen} is a compiler pipeline and runtime environment for standard C and Fortran code. The code is compiled into binaries containing CPU code and GPU kernels. At runtime the GPU kernels are automatically parallelized and JIT-compiled using runtime information. KernelGen is specialized on NVIDIA GPUs, other types of accelerators are not supported.

The Sambamba project~\cite{streit2012sambamba} aims at dynamically adapting programs to available resources at runtime. C/C++ code is compiled into an intermediate form, containing function definitions in sequential and automatically parallelized versions. This intermediate code is linked with a runtime environment into a fat binary. The runtime environment just-in-time compiles the intermediate code and dynamically adapts the execution by gathering information and running either the sequential or parallel version of functions per call. Similar to classic just-in-time compilation, the adaptation is limited to the target machine the program is started on.

\section{Design and  Implementation}
BAAR has a client-server architecture consisting of modules with well-defined tasks. The client runs the original program and provides proxy functions that optionally forward the function call for execution to the server. The server receives the code to be accelerated from the client, compiles the code for the target architecture, and provides access to the accelerated functions to the client through a remote procedure call interface.

\subsection{BAAR Server}
The server is responsible for providing an acceleration target for the client. It optimizes code it receives from the client for the accelerator and executes remote procedure calls. When started, the server starts listening for client connections. A connecting client sends its remote part, containing LLVM IR code to be offloaded. The server optimizes this code for the current accelerator available, which is the Intel Xeon Phi for our prototype. The Intel Xeon Phi accelerator card runs a dedicated Linux itself besides the operating system running on the host CPU. This enables us to run the server directly on the accelerator itself instead of on the host CPU, which saves the overhead of copying arguments from the host CPU to the accelerator card when executing function calls on it. The Intel Xeon Phi provides up to 61 cores, each able to run four threads in parallel and provides 512bit vector instructions. Therefore, the optimizations performed on the server side focuses on parallelization and vectorization.

Parallelization is performed before vectorization to obtain coarse-grained subfunctions which can be distributed among many threads and potentially contain parallelism which can further be exploited by vectorization. Automatic parallelization performed by the BAAR Server is based on the LLVM subproject Polly~\cite{grosser2011polly}, which provides an interface to polyhedral optimizations using CLooG and isl to the LLVM world. Polyhedral optimizations operate on Static Control Parts (SCoP), maximal sets of consecutive statements where loop bounds and conditionals are affine functions of the surrounding loop iterators and the parameters~\cite{Griebl98codegeneration}.  Polly can detect SCoPs in the offloaded LLVM IR code, models them in polyhedral representation and generates code with more parallelism exposed. It provides an OpenMP code generation backend which transforms loop bodies in detected SCoPs into subfunctions and inserts OpenMP library calls in the function to distribute execution of iterations among several threads. The resulting code can be further optimized using vectorization.

LLVM supplies powerful vectorizers of which BAAR uses the Loop Vectorizer and the Superworld-Level Parallelism (SLP) Vectorizer. The Loop Vectorizer supports powerful features, e.g., inserting runtime alias checks or vectorizing mixed types, which allow complex loop vectorization to be performed by the server. The SLP Vectorizer focuses on combining similar independent instructions into vector instructions. As vectorization relies on target-specific information and the Intel Xeon Phi is not supported as a target by LLVM, the x86 target is used to perform vectorization while forcing the vectorizers to emit 512bit vector instructions. This way vectorized LLVM IR code suitable for the Xeon Phi architecture is generated.

\begin{figure} 
	\begin{center}
	\includegraphics[width=\linewidth]{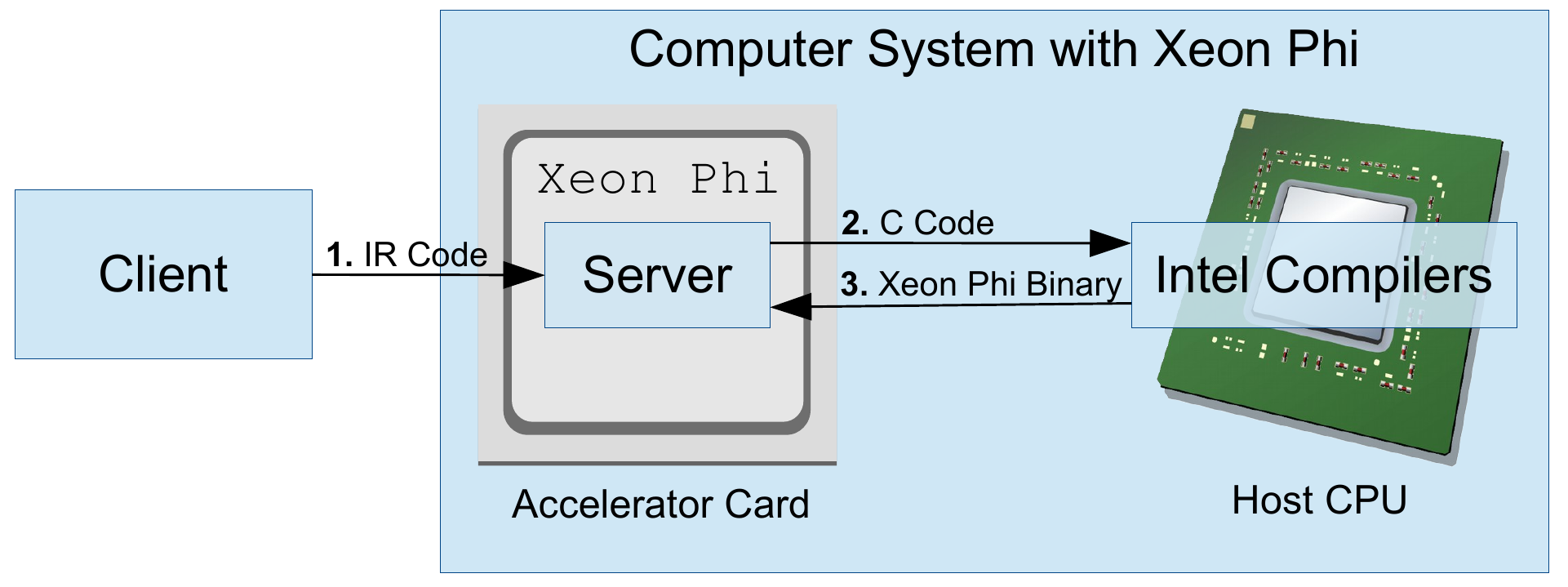}
	\end{center}
	\small{
	\begin{enumerate}[1.]
		\item Client sends IR code (remote part) to server
		\item Server on Xeon Phi transforms IR code into C code and sends it to host system
		\item Host system compiles C code to Xeon Phi binary using Intel Compilers, binary is sent to server
	\end{enumerate}}
	\caption{Mechanism to obtain a native Xeon Phi binary from LLVM IR}\vspace{-1.3em}
	\label{fig:c_detour}
\end{figure}

As the Intel Xeon Phi is currently not supported by LLVM, there exists no standard way to obtain Xeon Phi binaries from LLVM IR code. Furthermore, the Intel Compilers are currently the only compilers which are able to generate binaries for the Intel Xeon Phi which make use of its vector units. Unfortunately, the Intel Compilers are unavailable on the Xeon Phi itself. Therefore, the BAAR Server on the Xeon Phi has to transform the LLVM IR code into a representation accepted by the Intel Compilers and the use a compiler on the host CPU to generate binaries. The resulting design of the prototype uses the LLVM C backend to transform the LLVM IR code into C code. This backend was formerly part of LLVM but has been deprecated due to a lack of maintenance. Fortunately, Intel revived the C backend in the ispc project and extended it to support vector instructions using intrinsics for several architectures, including the Intel Xeon Phi. In the BAAR Server, the C code obtained from the optimized remote part is copied to the host CPU and the Intel Compilers are initiated to obtain an Intel Xeon Phi binary in the form of a shared object file. This shared object file is then copied back to the BAAR Server and dynamically linked with the server application. The whole process is depicted in Figure~\ref{fig:c_detour}. For performing function calls on behalf of the client, the type information available in the LLVM IR has to be combined with the native functions available as symbols in the shared object, accessible through void pointers. For this purpose, we use the foreign function interface library (libffi).

Once the server has optimized the remote part of a client and initialized the resulting binary, the client can call the functions contained in the remote part on the server to achieve accelerated execution.

\subsection{BAAR Client}
The client is responsible for providing an environment to execute and analyze programs present in LLVM IR. A central part of it is the LLVM Execution Engine, a library for running LLVM IR code with just-in-time compilation features. The Execution Engine enables the client to execute a program as well as  analyze, alter and recompile parts of it in parallel.

Identifying suitable function calls to offload to the accelerator provided by the server is performed in three steps: profiling, scoring and deciding per call at runtime. The profiling step uses LLVM Passes to estimate the maximum basic block frequency (estimated number of executions) for every function. A function containing a basic block with a high frequency is a promising candidate for acceleration with the BAAR Server, as such a basic block indicates the existence of a loop with high iteration counts, which can potentially be sped up greatly when parallelization and vectorization are applied. Functions containing a basic block with a frequency exceeding a user-configurable threshold are gathered in a candidate set to be further analyzed.

The second step is to score every function from the candidate set for suitability for the accelerator. The score gives an abstract value of suitability, by default functions with a score higher than zero are exported into the remote part. For the BAAR prototype, the only available accelerator is the Intel Xeon Phi which requires over one hundred threads to achieve its full performance. Therefore, the score is based on the number of operations which can be parallelized during optimization on the BAAR Server. For the automatic parallelization on the server side, the existence of SCoPs is crucial. Thus, the first step in the scoring process of a function is to try to detect SCoPs using Polly's \texttt{SCoPDetection} Pass. If no SCoP is detected, the candidate function is discarded and not considered anymore for acceleration. A function containing at least one SCoP is further analyzed, the analysis iterates over every detected SCoP and in every detected SCoP over every contained loop nest. For every loop nest, the total numbers of floating point and integer operations are determined separately. These two counts can be weighted by configurable factors, the standard weight is one for both counts. The weighted counts are added and the result is multiplied to the innermost basic block frequency of the current loop nest to give a weighted estimate of total floating point and integer operations performed when the current loop is executed. To determine the innermost basic block frequency, the LLVM Passes \texttt{LoopInfo} and \texttt{ScalarEvolution} are utilized. The weighted estimate of total floating point and integer operations defines the loop nest's score. The SCoP's score is the sum of loop nest scores, and the function's score is the sum of SCoP scores:
\small\begin{equation}
\begin{split}
\text{score}_\text{loop} &= (c_\text{IOPs} \cdot \text{IOPs}_\text{loop} + c_\text{FLOPs} \cdot \text{FLOPs}_\text{loop}) \cdot \text{innerBBFreq}_\text{loop} \\ 
\text{score}_\text{function} &= \sum\limits_{\text{SCoP}\in\text{function}} \ \sum\limits_{\text{loop}\in\text{SCoP}}\text{score}_\text{loop} \label{eqn:score}
\end{split}
\end{equation}
\normalsize
When the scoring finishes, the remote part is complete and is sent to the server. Note that the whole program analysis and building of the remote part happens in parallel to the execution of the original program, transparently for the user.

After the server signals its readiness to perform remote calls on the functions contained in the remote part, the client transforms the original program into the local part. This entails the preparation of the third step of identifying suitable function calls for offloading: insertion of the per-call runtime decision of whether to execute the call remotely or locally. Functions in the remote part are considered suitable for acceleration, however there may be calls to a suitable function for which the time taken to transfer the arguments is longer than the time saved by acceleration. To run these calls locally while offloading the others, a per-call runtime decision based on score and arguments size is introduced. For every function contained in the remote part, the function definition is altered to begin with the calculation of the total number of bytes needed to be transferred when executing the call remotely. The total number of bytes transferred for our simple communication protocol is the size of the arguments plus the size of the result. Currently, this means that the size of array arguments is counted twice, as arrays are always transferred to and from the server as a whole. After the calculation of the total number of bytes transferred, the actual runtime decision is inserted. It compares the fraction of the function's score over the total number of bytes transferred to a system-dependent, user-configurable constant $c$. If the fraction is bigger than $c$, the function call is executed remotely. Otherwise, the original function body is executed locally.

\begin{figure} 
	\begin{center}
	\includegraphics[width=\linewidth]{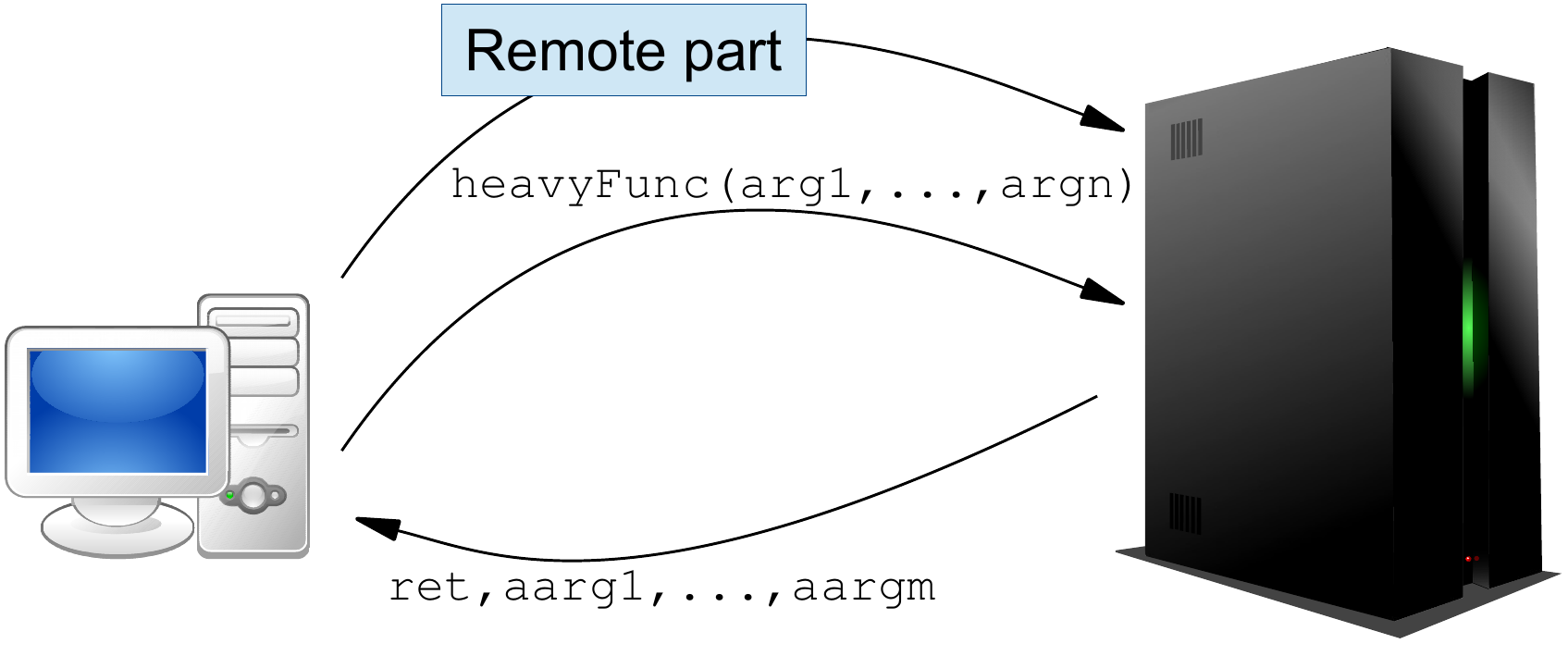}
	\end{center}
	\caption{The client sends the remote part to the server. Once the server finishes initializing acceleration, the client can perform function calls remotely on it. The results and arguments potentially altered during execution are sent back to the client.}\vspace{-1.3em}
	\label{fig:remote_exe}
\end{figure}

When executing a function call remotely, the call information have to be transferred to the server. Currently, the function signature and arguments are simply marshalled into a string representation and sent to the server over the TCP/IP socket. The server unmarshalls the information and executes the call on the Xeon Phi. Once the call finishes, the return value and arguments which could potentially have changed are sent to the client, again marshalling the values into a string representation. A simplified view on this process is given in Figure~\ref{fig:remote_exe}.

\section{Evaluation}
We have evaluated the feasibility and performance of BAAR using two different stencil applications (2D Jacobi and 2D FDTD), which are taken from the Polybench~\cite{polybench} benchmark suite. The applications are executed using BAAR with several problem sizes and iteration counts. The BAAR Client is executed on a dual Intel Xeon E5-2670 machine with 8 cores running at 2.6 GHz and 64 GB of RAM. The same system hosting the client was also used to measure execution times as a basis for evaluating possible speedups. The server is run on an Intel Xeon Phi 5110P with 60 cores at 1.053 GHz each and 8 GB of RAM. The operating system is Scientific Linux 6.4 with the Intel Many Core Platform Stack in version 2.1.6720. The stencil codes are written in C code which is compiled using LLVM 3.4 into LLVM IR to be executed with BAAR. As a baseline for evaluating the quality of BAAR, we compare the achieved performance of BAAR with the performance of the same C code compiled using the Intel Compiler in version 14.0.0 using optimization level O2 (optimization for speed) running natively on the Xeon E5. The setup for this experimental evaluation reflects the choice a potential user has: either executing his existing binary generated with a standard compiler or execution using BAAR.

\subsection{Program Analysis and Acceleration Initialization}
Analysis and initialization are the investments made to yield speedups later on. The acceleration initialization consists of optimizing the remote part, initializing the acceleration target, transforming the original program to the local part as well as communication between client and server. Optimization and accelerator initialization are not time-critical as these are tasks for the server only and the client concurrently makes progress on the original program without any overhead. Analysis and program transformation are time-critical as they are performed on the client and must not impair the concurrent program execution. 

\begin{figure} 
	\begin{center}
	\includegraphics[width=\linewidth]{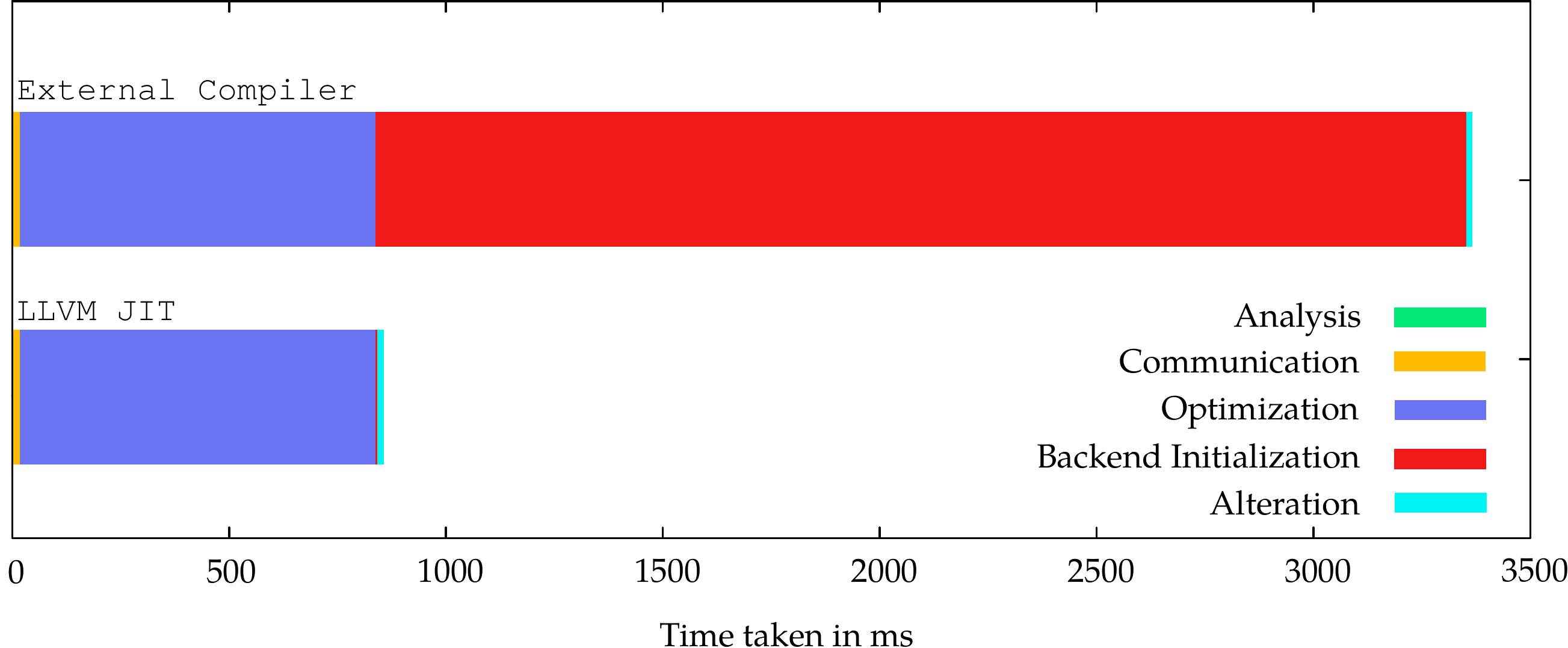}
	\end{center}
	\caption{Time taken for program analysis and acceleration initialization}
	\label{fig:acc_init}
\end{figure}

Figure~\ref{fig:acc_init} shows the times taken for program analysis and acceleration initialization on an example program containing the Jacobi 2D stencil. \texttt{External Compiler} denotes the current BAAR implementation utilizing the host compiler to obtain Xeon Phi binaries, therefore the overall time taken is dominated by the target initialization on the server side. \texttt{LLVM JIT} shows the same time measures, but the target initialization is exchanged with a time measure performed on the Intel Xeon E5 with the native x86 JIT backend. This shows that we can expect that once the native Xeon Phi backend becomes available, the total time taken for analysis and acceleration initialization of 853 ms is almost solely spent on optimization ($>$ 96\%), a non-time-critical server task. The client only spends 27 ms (0.032\% of \texttt{LLVM EE}, 0.008\% of \texttt{ExtComp}) in the analysis and initialization phases. For long running programs like stencil computations this overhead is quickly amortized even if only minor speedups are achieved. Further, the compilation for Xeon Phi can be considered as a one-time overhead applying only once for each distinct remote part, when previously generated binaries are cached.

\subsection{Jacobi 2D}
\begin{figure} 
	\begin{center}
	\includegraphics[width=\linewidth]{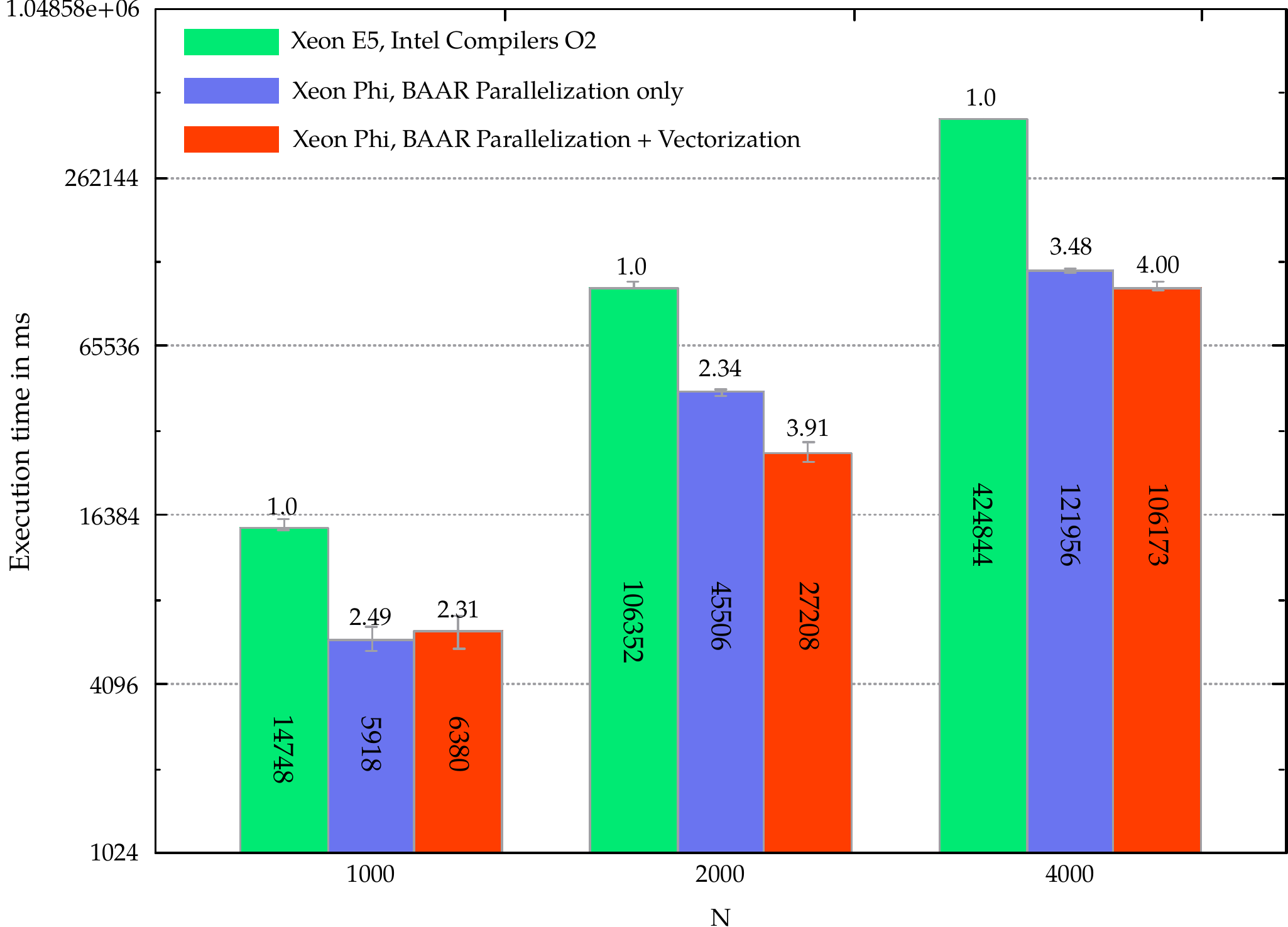}
	\end{center}
	\caption{Raw execution times for Jacobi 2D, 10000 iterations (log scale)}\vspace{-1.3em}
	\label{fig:jacobi2d_S10000}
\end{figure}

Figure~\ref{fig:jacobi2d_S10000} compares performance results of executing the Jacobi 2D stencil optimized using the Intel Compilers on the Xeon E5 and of executing the stencil using BAAR's automatic optimizations at runtime on the Xeon Phi. The graph only considers the raw execution times, i.e., the time taken to finish the function call on the respective target without considering overheads imposed by remote execution (un-/marshalling). The number of iterations is set to 10000, executing on an array of size N$^2$ with N$\in\{1000, 2000, 4000\}$. At N = 2000, the speedup of the parallelized and vectorized code on the Xeon Phi over the heavily optimized code on the Xeon E5 is 3.91$\times$. The speedup rises to 4$\times$ in these measurements when N is doubled to 4000. This means the raw execution time is over 5 minutes less on the Xeon Phi (1 min 46 s) compared to the Xeon E5 (7 min 5 s). The measurements also show that the vectorization profits from data sizes bigger than N = 1000. Setting N = 2000, the speedup with vectorization and parallelization compared to only parallelization is 1.67$\times$. However, with N = 4000 the speedup drops to 1.15$\times$. The exact reasons for this aberration have to be clarified in future research. It could be an issue of data alignment, which was not taken in consideration in the current state of BAAR. Further, the C backend from Intel's ispc we are reusing in this work supports several parameters. Potentially, the parameters chosen for BAAR are not optimal.

The speedup of up to 4$\times$ achieved completely transparently without any developer-provided directives on a Xeon Phi over an optimized parallelized and vectorized execution on a Xeon E5 is a promising result for the practicality of BAAR. Especially when considering that the Xeon E5 is a high end processor, which potential clients may not have available.

\begin{table}[htb]
\begin{center}
\begin{tabular}[c]{|r|r|r|r|}
\hline
	\diagbox{Value}{N} & 1000 & 2000 & 4000 \\ \hline
	Average & 12406 ms & 49945 ms & 195859 ms \\ \hline
	Minimum & 11562 ms & 47984 ms & 193814 ms \\ \hline
	Maximum & 13165 ms & 52912 ms & 201101 ms \\ \hline \hline
	Speedup	& 1.19 & 2.13 & 2.17 \\ \hline
	Spent in raw & 51.43 \% & 54.48 \% & 54.21 \% \\
	execution (avg.)& (6380 ms) & (27208 ms) & (106173 ms) \\ \hline
\end{tabular}
\end{center}
\caption{Time taken for \texttt{callAcc} calling \texttt{jacobi\_2d} with STEPS = 10000 and full optimization}\vspace{-1.3em}
\label{tbl:callacc}
\end{table}

However, so far we have only considered the raw execution time. To evaluate the system-level impact of BAAR, the overhead imposed by marshalling and unmarshalling the call for the remote procedure call has to be taken into account. Results that evaluate this scenario are presented in Table~\ref{tbl:callacc} which is based on the same set of measurements as Figure~\ref{fig:jacobi2d_S10000} but considers all  overheads for calling Jacobi 2D remotely instead of just the raw execution time. The data shows that from the previously calculated speedup of 4$\times$, when just comparing raw execution times with N = 4000 and STEPS = 10000, a speedup of 2.17$\times$ remains when considering the full remote call overheads. When increasing STEPS for a fixed N, the share of the communication of the time spent decreases so that the speedup of the complete call is closer to the raw speedup. It is evident however, that the basic socket-based communication mechanism currently implemented in BAAR heavily impairs the execution of remote calls. Using sockets as a means of communication is appealing because they are universally usable. For BAAR however, more suitable alternatives exist. A promising alternative for future research is the \textit{Message Passing Interface} (MPI), a standard for exchanging messages in distributed systems. It supports more sophisticated mechanisms to transfer values than transforming them into a string representation, as it is currently done in our proof of concept. MPI allows a much more efficient communication over, e.g., PCI Express between a client on the CPU and a server running on a host-internal accelerator card, as well as over network when running client and server on separate hosts.

\subsection{FDTD 2D}
In addition to the Jacobi stencil we evaluate a finite-difference time-domain (FDTD) stencil~\cite{Sulliv2013} application in two dimensions, which is an algorithm that is used for numerically solving Maxwell's equations. While Jacobi 2D was accelerated by BAAR without any developer-provided directives, BAAR has to be hinted to ignore the (false positive) results from aliasing analysis in order to identify the SCoPs in FDTD 2D. Otherwise, parallelization using Polly would have failed.

\begin{figure} 
	\begin{center}
	\includegraphics[width=\linewidth]{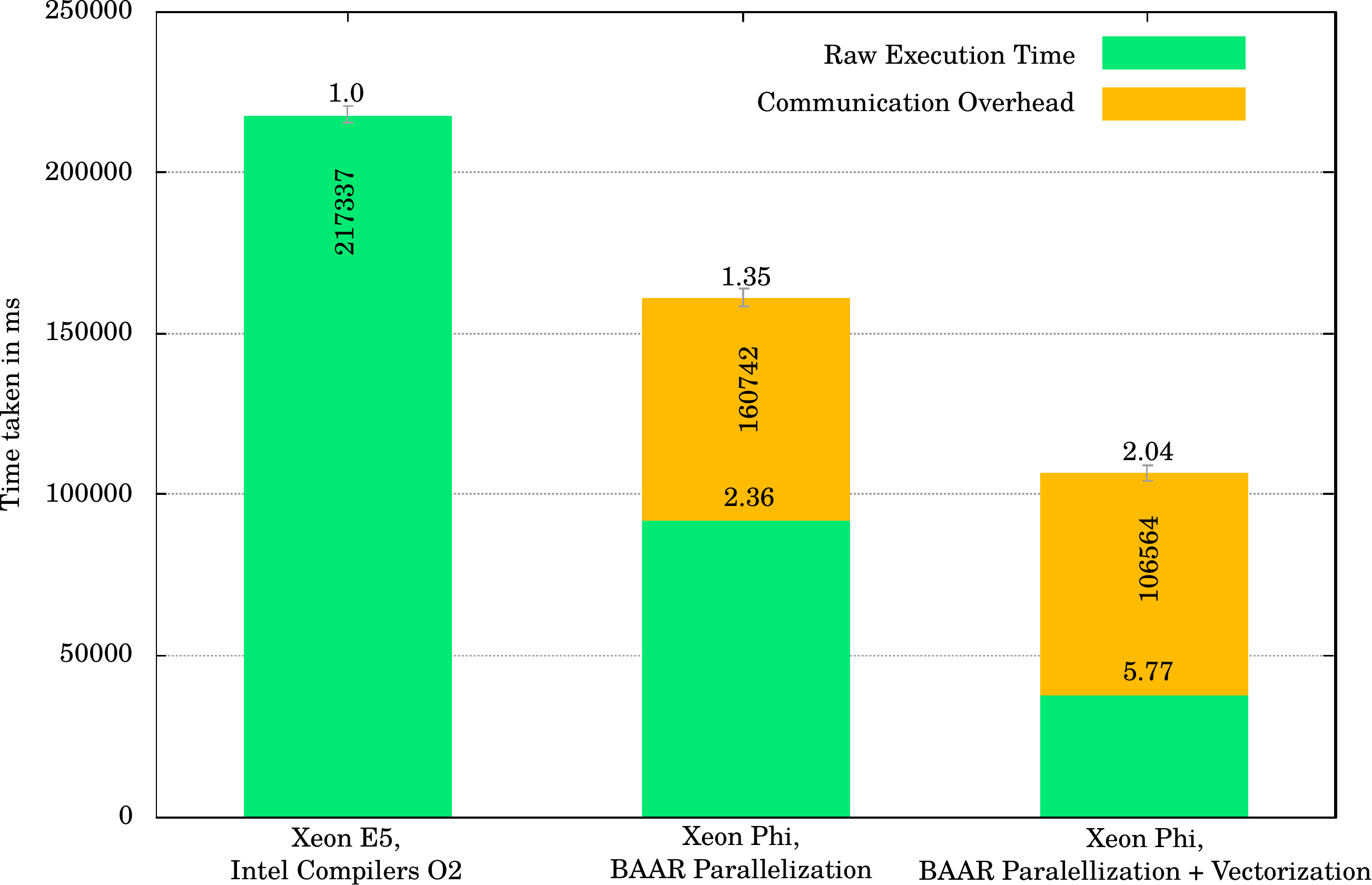}
	\end{center}
	\caption{Execution times for \texttt{fdtd\_2d} with STEPS = 10000, N = 2000}\vspace{-1.3em}
	\label{fig:fdtd_2d_stacked}
\end{figure}

Figure~\ref{fig:fdtd_2d_stacked} shows average execution times for FDTD 2D with STEPS = 10000 and N = 2000 over ten runs and the achieved speedups compared to execution of code compiled with the Intel C++ Compiler with optimization level O2 on an Intel Xeon E5-2670. Considering only the raw execution times measured by executing the code with BAAR, the speedups are 2.36$\times$ when using only parallelization and 5.77$\times$ for parallelization and vectorization. When comparing these results to the measurements taken for Jacobi 2D, the speedup when using only parallelization is very similar. Enabling vectorization in addition to parallelization gives the execution on BAAR an additional speedup of 2.44$\times$, resulting in a speedup of 5.77$\times$ in total.  In comparison to the simpler Jacobi 2D stencil, the effect of vectorization is much higher. This stresses that future research should improve the parameters used for vectorization in BAAR and investigate the requirements for more reliable results.

When considering the full time taken for executing FDTD 2D remotely, the raw speedups of 2.36$\times$ and 5.77$\times$ are decreased to 1.35$\times$ and 2.04$\times$ overall speedup, respectively. This underlines that the simple communication mechanism impairs the execution and should be replaced.

\section{Discussion and Future Work}
Our work shows that transparent acceleration by offloading computational hotspots of binary applications to a many-core accelerator is feasible and, given suitable applications, yields significant speedups over the executing the original code on the CPU. The proposed approach is not restricted to a scientific computing scenario but is applicable for any situation where an accelerator is likely to significantly outperform the client CPU. Our case study demonstrates that even for powerful client CPUs performance can be gained when the compilers are not able to exploit the full parallelism of the client CPU, for example, because of limited alias analysis or vectorization capabilities. But we also envision benefits of our approach for embedded computing systems that have a high computational demand that cannot be satisfied by the CPU. Given the client-server architecture of BAAR, the approach is also suitable for sharing an accelerator between many clients that sporadically offload computational hotspots.

For the future, we plan to expand our work on BAAR in two main directions, making BAAR more widely applicable and improving the performance of the framework.
To make BAAR more widely applicable, several problems require further research:
\begin{itemize}
	\item The maturity and limitations of decompilation techniques\cite{dagger,Anand:2013:CIR:2465351.2465380,Chipounov2011} from binary to LLVM IR need to be inspected systematically with respect to the optimizations performed in BAAR.
	\item It would be very interesting to evaluate the performance of BAAR with different combinations of client systems and server system. E.g., the BAAR Client is easily portable to the ARM architecture. This leads to an interesting setup consisting of several BAAR Clients running on low-power ARM-based systems sharing an Intel Xeon Phi accelerator for their calculations provided by the BAAR Server.
	\item The server should be extended to provide several targets to the clients, not only one target throughout its whole lifetime. It would increase the potential for increased performance to have several targets with different characteristics the client can choose from, e.g., Xeon Phi, FPGA, GPU, etc.. Additionally, it would be interesting to investigate in criteria other than performance. Maybe the BAAR Server can provide a target to the client which can perform a certain calculation much more power efficient, during calculation the client can standby and save power. It is also thinkable to switch the target when requirements change.
	\item The constant $c$ which is compared to $\frac{\text{function score}}{\text{bytes to transfer}}$ of a certain function call to make the decision whether it should be run remotely or locally, is currently set by the user at server start or defaults to zero. $c$  should automatically be determined with simple benchmarks and could additionally be self-adapting.
	\item Currently, only the size of the arguments of a certain call are used as runtime information. Especially alias analysis could profit from utilizing runtime information and increase the applicability of BAAR by enabling more complex functions to be identified as suitable for remote execution.
\end{itemize}

For further improving the performance of BAAR, we plan to investigate the following directions:
\begin{itemize}
	\item The evaluation has shown that the communication mechanism is currently the weak point of BAAR when trying to achieve large speedups. A communication mechanism between client and server which utilizes MPI and, if available, an Infiniband network, could significantly improve the performance of an offloaded call. Especially argument marshalling would profit from a more sophisticated mechanism than currently implemented.
	\item When having arrays as arguments, they are always transferred to and from the server as a whole for every call. Ideally, a more intelligent mechanism would only copy elements to the server which are used by the code executed on it and copy back elements which were altered. Additionally, it could make sense to cache arrays for subsequent calls.
	\item At this point, a function which is called before the acceleration is initialized is executed locally by the client, even when the initialization finishes while the call is still executing. This situation could be improved by introducing \textit{on-stack-replacement}. With on-stack-replacement, the running function call would be suspended once the acceleration finishes, the remote execution of this call for the current state of execution would be setup and the call would be continued by running it remotely on the accelerator. Especially long running function calls could greatly profit from this extension to BAAR.
	\item On a more technical level, BAAR would greatly profit from a native LLVM backend for the Intel Xeon Phi. It would enable us to drop the detour over C code and improve the quality of vectorization.
\end{itemize}

\section{Availability}
BAAR is open-source software available at \url{https://github.com/pc2/baar} under the MIT license.

\section{Conclusion}
This paper introduced BAAR, our approach of tackling the problem of enabling existent software to automatically utilize accelerators. BAAR is capable of automatically detecting functions suitable for acceleration, offloading them as well as automatically parallelizing and vectorizing them. Whether to utilize the accelerated remote execution is decided at runtime per function call. The whole process is transparent to the user. The evaluation could show BAAR's practicality by achieving a speedup of up to 4$\times$ without any developer-provided hints, when comparing execution of a real-life example compiled with the Intel C++ Compiler at optimization level O2 on a dual CPU 8-core Intel Xeon E5-2670 system to execution using BAAR utilizing an Intel Xeon Phi 5110P accelerator card. With developer-provided hints, even a speedup of 5.77$\times$ could be achieved. This points out the performance possible when alias analysis is improved.

\section*{Acknowledgment}
This work was partially supported by the German Research Foundation (DFG) within the Collaborative Research Centre ``On-The-Fly Computing'' (SFB 901) and the European Union Seventh Framework Programme under grant agreement no.~610996 (SAVE).

\bibliographystyle{IEEEtran}
\bibliography{damschen14_adapt}
\end{document}